\documentstyle[preprint,prl,aps]{revtex}
\begin{document}


\title{Shear viscosity and shear thinning in two-dimensional Yukawa
liquids}

\author{Z. Donk\'o$^1$, J. Goree$^2$, P. Hartmann$^1$, and K. Kutasi$^1$}

\address{$^1 $Research Institute for Solid State Physics and Optics of
the Hungarian Academy of Sciences,\\ H-1525 Budapest, P.O. Box 49,
Hungary}

\address{$^2 $Department of Physics and Astronomy,
The University of Iowa, Iowa City, Iowa 52242, USA}

\date{\today}

\maketitle

\begin{abstract}

A two-dimensional Yukawa liquid is studied using two different
nonequilibrium molecular dynamics simulation methods. Shear
viscosity values in the limit of small shear rates are reported
for a wide range of Coulomb coupling parameter and screening
length. At high shear rates it is demonstrated that this liquid
exhibits shear thinning, i.e., the viscosity $\eta$ diminishes
with increasing shear rate. It is expected that two-dimensional
dusty plasmas will exhibit this effect.

\end{abstract}



Many-particle systems characterized by the Yukawa potential
include a variety of physical systems, e.g. dusty plasmas, charged
colloids, astrophysical objects and high energy density matter.
The Yukawa potential $\phi(r) \propto Q \exp(- r / \lambda_{\rm
D}) / r$ models a Coulomb repulsion that is exponentially
suppressed with a screening length $\lambda_{\rm D}$. Yukawa
systems behave like a liquid when the temperature exceeds a
melting point which depends on $Q$, $\lambda_D$, and particle
spacing, e.g. \cite{khrapak,H05}.

Transport parameters of Yukawa systems -- the diffusion coefficient
\cite{OH00}, the shear viscosity \cite{SM01,SH02,SC02FM03} and the thermal
conductivity \cite{SC02FM03,DH04} -- have mainly been calculated for 3D
systems, but there is now an increasing interest in 2D settings. For example,
in dusty plasma experiments, charged microspheres  suspended as a monolayer in
a gas discharge make a 2D Yukawa system. By creating a shear flow in such a
particle suspension, the viscosity was measured in recent experiments using 2D
suspensions \cite{NG04} and quasi-2D suspensions consisting of a few monolayers
of charged microspheres \cite{thinning}. The transport properties of such
ultrathin liquids are also of interest as macroscopic analogs of molecular flow
in nanoscience applications \cite{meso}.

Transport coefficients are meaningful if they are part of a valid
``constitutive relation'' between the gradients of local variables
and fluxes. For shear viscosity $\eta$, the constitutive relation
$j_y = -\eta [dv_x(y) / dy]$ relates a momentum flux $j_y$ to the
velocity gradient $dv_x(y)/dy$, which is also termed the shear
rate. In a non-Newtonian fluid, $\eta$  may vary with the velocity
gradient, whereas in Newtonian fluids it does not. In particular,
if $\eta$ diminishes as shear is increased, the fluid is said to
exhibit ``shear thinning''. This occurs in simple liquids
\cite{evans}, as well as in complex mixtures such as foams,
micelles, slurries, pastes, gels, polymer solutions, and granular
flows \cite{banding}. Recently, experimenters have claimed to
observe shear thinning in dusty plasma liquids \cite{thinning}.
These reports motivate our simulations to search for the presence
of shear thinning in 2D Yukawa liquids.

Subsequent to the experimental measurement of viscosity in a 2D
dusty plasma \cite{NG04}, a 2D molecular dynamics simulation was
used to obtain the shear viscosity from the Green-Kubo relations
\cite{LG05}. In this Letter we will go beyond the results of
Ref.~\cite{LG05}, which were performed for equilibrium conditions,
by using here nonequilibrium simulations to search for
non-Newtonian behavior under conditions of a high shear rate. We
will also compute the viscosity over a wider range of $\Gamma$ and
$\kappa$.

Our simulations use a rectangular cell with edge lengths $L_x$ and
$L_y$ and periodic boundary conditions. The number of particles is
between $N$ = 990 and 7040. The system is characterized by
dimensionless parameters $\Gamma = Q^2 / 4 \pi \varepsilon_0 a k_B
T$ and $\kappa = a / \lambda_{\rm D}$, where $a=(1/n \pi)^{1/2}$
is the Wigner-Seitz radius, with $n$ being the areal density.
Additional parameters include the thermal velocity  $v_0 = (2
k_{\rm B} T /m)^{1/2}$ and the 2D analog of the plasma frequency
$\omega_{\rm p} = (Q^2/2 \pi \varepsilon_0 m a^3)^{1/2}$, the
shear rate $\gamma = dv_x/dy$, and its normalized value
$\overline{\gamma} = (dv_x/dy) (a/v_0)$. Two types of molecular
dynamics techniques are applied for the studies of the shear
viscosity.

Method 1 reverses the cause-and-effect picture customarily used in
nonequilibrium molecular dynamics: the effect, the momentum flux,
is imposed, and the cause, the velocity gradient (shear rate) is
measured in the simulation \cite{FMP}. Momentum in the liquid is
introduced in a pair of narrow slabs A and B, which are centered
at $y=L_y/4$ and $3L_y/4$, respectively. At regular time intervals
$\tau$ we identify the particles in slabs A and B having the
highest $|v_x|$ in the positive and negative directions,
respectively. We then instantaneously exchange the $v_x$ velocity
component of these two particles without moving the particles.
This artificial transfer of momentum between slabs A and B (which
is accomplished without changing the system energy) produces a
velocity profile $v_x(y)$, the slope of which can be controlled by
the frequency of the momentum exchange steps. The equations of
motion
\begin{equation} \label{eq:motion}
    \frac{d {\bf r}_i}{dt}=\frac{{\bf p}_i}{m}, \hspace{1cm} \frac{d {\bf p}_i}{dt}={\bf F}_i,
\end{equation}
where ${\bf r} = (x,y), {\bf p} = (p_x,p_y)$ are the positions and
the momenta of particles, $m$ is their mass, and ${\bf F}_i$ is
the force acting on particle $i$, are integrated by the velocity
Verlet algorithm.

Method 2 simulates a planar Couette flow, which is established by
the Lees-Edwards periodic boundary conditions which result in a
homogeneous streaming flow field in the simulation box: $\langle
v_x \rangle = \gamma (y - L_y/2)$, where $\langle \rangle$ denotes
a time average. The system is described by the Gaussian
thermostated SLLOD equations of motion \cite{book}:
\begin{eqnarray} \label{eq:sllodmotion}
    \frac{d {\bf r}_i}{dt}=\frac{{\bf \tilde{p}}_i}{m} + \gamma y_i {\bf \hat{x}},
\nonumber \\
    \frac{d {\bf \tilde{p}}_i}{dt}={\bf F}_i -
    \gamma \tilde{p}_{yi} {\bf \hat{x}} - \alpha{\bf \tilde{p}}_i,
\end{eqnarray}
where ${\bf \tilde{p}} = (\tilde{p}_x,\tilde{p}_y)$ is the
\textit{peculiar} momentum of particles, ${\bf \hat{x}}$ is the
unit vector in the $x$ direction, and $\alpha$ is the Gaussian
thermostatting. multiplier. The above set of equations is solved
using an operator splitting technique \cite{osa2005}.

In contrast to method 1, method 2 results in a homogeneous shear
field and a constant temperature within the whole simulation box.
Thus arbitrarily high shear rates may be established without the
need of considering any effects of temperature gradients on the
viscosity.

In both methods the pairwise Yukawa interparticle forces are summed
over a $\kappa$-dependent cutoff radius, using the chaining mesh
technique. (The force due to particles at the cutoff radius is
$\approx 10^{-5}$ smaller compared to that due to the nearest
neighbors.) Both methods neglect any neutral gas drag, which has
been observed to alter the velocity profiles in experiments
\cite{NG04}, i.e. they model an atomic system where momentum
transfer is dominated by Coulomb collisions \cite{khrapak}.

Method 1 has the advantage that is resembles more closely the
experimental conditions, although the procedure for applying shear
in the simulation involves no introduction or removal of energy
from the system. In the experiment \cite{NG04} shear is applied
via an external introduction of both momentum and energy in a
boundary slab while energy is simultaneously removed elsewhere by
frictional dissipation. Method 2 represents a well-established
technique for measurement of viscosity at arbitrary steady, as
well as temporally varying shear rates. Although it has little
connection to the conditions found in the experiment, it has been
demonstrated to be an efficient technique to investigate shear
thinning \cite{evans,book}. Thus we apply this method for the
studies of this latter effect.

Near equilibrium (small $\gamma$) shear viscosity values have been
obtain using both techniques. In method 1 this is done at the
lowest practical shear rate, where $d v_x / dy$ is uniform between
slabs A and B. We calculate $\eta_{\rm eq}$ from
\begin{equation}
|j_y| = \eta_{\rm eq} dv_x(y) / dy = \Delta p / 2 t_{\rm sim} L_y,
\end{equation}
where $\Delta p$ is the total $x$-directional momentum exchanged between slabs
A and B during the simulation time $t_{\rm sim}$ \cite{FMP}. In method 2 the
off-diagonal element of the pressure tensor is measured during the course of
the simulation:
\begin{equation}\label{eq:pressuretensor}
    P^{xy}(t) = \sum_{i=1}^N \Bigl[ m v_{ix} v_{iy} + \sum_{j>i}^N
    \frac{x_{ij} y_{ij}}{r_{ij}} \frac{d}{dr_{ij}} \phi(r_{ij})
    \Bigr],
\end{equation}
where ${\bf r}_{ij} = {\bf r}_i - {\bf r}_j = (x_{ij},y_{ij})$, and the shear
viscosity is obtained as
\begin{equation}\label{eq:sllodeta}
  \eta = \lim_{t \rightarrow \infty} \langle P^{xy}(t) \rangle / \gamma.
\end{equation}

In method 1, the spatial profiles for temperature and velocity,
Fig.~\ref{fig:examples}, develop self-consistently in response to
the perturbation applied by introducing momentum in slabs A and B.
We use method 1 only for small perturbations, so that the velocity
profile has a linear gradient and the temperature is isotropic,
with $T_x = T_y$, where $T_{x,y} = (m/N_j k_{\rm B})
\sum_{i=1}^{N_j} \bigl< [ v_{ix,y}(t) - \overline{v_{jx,y}} ]^2
\bigr>$. The index $i$ runs over the $N_j$ particles in slab j. We
verified that $\overline{v_{jy}}$ is negligibly small.

Obtaining reliable results for $\eta$ at small $\gamma$ requires a
simulation duration of typically $\omega_p t \sim$ 10$^4$ --
10$^5$ for both methods. The required time step is smallest, and
the simulations are most costly, at low $\Gamma$. In method 1,
system size effects are expected to appear when (i) particles
traverse the simulation box without significant interaction with
the others, or (ii) the compressional sound wave transits the box
in a shorter time than the decay time $t_c$ of the velocity
autocorrelation function. For (i), for our most demanding
condition (small size $N$ = 990 and high temperature $\Gamma$ = 1)
a particle moving at the thermal velocity would transit the cell
in a time $\omega_{\rm p} t \approx $ 57 if it were undeflected by
collisions. We find that the decay time is short enough,
$\omega_{\rm p} t_c \sim $ 5 -- 10, even for the smallest $\Gamma$
values of interest. Thus we expect no ``ballistic'' trajectories
across the entire simulation box. For (ii), the sound speed
\cite{PRL2004} at $\kappa$ = 1 is $v = d\omega / dk \sim a
\omega_{\rm p}$, and the wave's transit time $\Delta t$ across a
box with length $L_y$ is $\omega_{\rm p} \Delta t \sim L_y / a $ =
57 for our most demanding case, $N$ = 990 particles. Thus, we find
both criteria fulfilled for a ``sufficiently large'' system.
Method 2 is known to produce accurate results even for small
number of particles simulated \cite{book}. We verified that the
results obtained from both methods did not depend significantly on
$N$.

Figures~\ref{fig:sllod}(a) and (b) illustrate particle
trajectories in simulations based on method 2, for conditions
$\Gamma$ = 10, $\kappa$ = 1 at a shear rate $\overline{\gamma}$ =
0.2, and for $\Gamma$ = 100, $\kappa$ = 1, $\overline{\gamma}$ =
0.05, respectively.

Our results for $\eta_{\rm eq}$ as a function of $\Gamma$, for
different values of $\kappa$ are plotted in
Fig.~\ref{fig:results}(a). We find a good agreement with the
earlier equilibrium MD simulation of Ref.~\cite{LG05}. In contrast
with most simple liquids, which have a viscosity that varies
monotonically with temperature, a prominent feature of the
viscosity of the present system is a minimum (e.g. at $\Gamma
\cong$ 20 for $\kappa$ = 1), which has been noted previously in
both OCP (one-component plasma) and Yukawa liquids. The shape of
the $\eta_{\rm eq}(\Gamma)$ curve can be explained by the
prevailing kinetic and potential contributions to the viscosity at
low and high values of $\Gamma$, respectively. The
near-equilibrium shear viscosity values obtained with method 2 for
$\kappa$ = 1 are also displayed Fig.~\ref{fig:results}(a). We find
an excellent agreement between the results of methods 1 and 2.

Similar to what was observed in \cite{SH02} for 3D Yukawa liquids,
we find that the near-equilibrium viscosity $\eta_{\rm{eq}}$ obeys
a scaling law as demonstrated in Fig.~\ref{fig:results}(b), where
viscosity has been normalized by $\eta_{\rm E}= m n \omega_{\rm E}
a^2$. The Einstein frequency $ \omega_{\rm E}$ depends on
$\kappa$, and we computed it from Eq.~(7) of Ref.~\cite{PRL2004}
using pair-correlation functions measured from our simulations.
The horizontal axis is a normalized temperature $T' = T_y/T_m =
\Gamma_m / \Gamma$, where $T_m$ and $\Gamma_m$ are melting-point
values reported in Ref.~\cite{H05}. Using these normalizations,
the data fall on the same curve, demonstrating the existence of a
scaling law for the 0.5 $\leq \kappa \leq$ 2.0 range of the
screening parameter. We note that for this purpose we found $
\omega_{\rm E}$ was more significant than $\omega_{\rm p}$. The
near-equilibrium viscosity is fit by an empirical form
$\eta_{\rm{eq}} / \eta_{\rm E} = aT' + b/T' + c$ with
coefficients: $a$ = 0.0093, $b$ = 0.78 and $c$ = 0.098.

A shear-thinning effect is revealed in Fig.~\ref{fig:thinning}(a), which shows
that $\eta$ diminishes significantly as the shear rate $\overline{\gamma}$ is
increased. In other two-dimensional systems the reduction in $\eta$, as
compared to the value at small shear, was observed to vary as the square root
of $\overline{\gamma}$ \cite{evans}. We find that this scaling also occurs for
the Yukawa system, as indicated by data that fall nearly on a straight line in
Fig.~\ref{fig:thinning}(a) for $\overline{\gamma} > 0.2$. At smaller shear
rates, $\overline{\gamma} < 0.2$, however, the shear thinning effect is less
profound and the liquid is more nearly Newtonian, especially for large
$\Gamma$. Results are shown for $\overline{\gamma} \gtrsim 0.01$, which we
found to be reliable, whereas at lower $\overline{\gamma}$ method 2 yielded
noisy data even for very long simulations.

Because viscosity arises from both kinetic and potential
contributions, Eq.(\ref{eq:pressuretensor}), we evaluate which of
these contributions is most responsible for the observed
shear-thinning effect in Fig.~\ref{fig:thinning}(b). Recall that for
equilibrium conditions, the kinetic term dominates for $\Gamma \ll$
20 and the potential term dominates for $\Gamma \gg$ 20. Here, we
find that for non-equilibrium conditions, as the shear rate
increases the reduction in viscosity is mostly due to a reduction of
the kinetic contribution at low $\Gamma$ and a reduction of the
potential contribution at high $\Gamma$. In other words, at extreme
values of $\Gamma$, it is the same term dominating the equilibrium
viscosity that also dominates the shear thinning effect. At an
intermediate value of $\Gamma$ = 20, however, where the equilibrium
viscosity has a minimum and the two equilibrium contributions are
comparable (for $\kappa$ = 1), we find that it is mostly a reduction
of the kinetic term that accounts for the observed shear thinning
effect.

In summary, we have calculated the shear viscosity coefficient of 2D Yukawa
liquids in a wide domain of parameters $\Gamma$ and $\kappa$ using two
different molecular dynamics approaches. The small shear rate calculations
confirmed that the two techniques used are consistent and yielded $\eta$ values
in fair agreement with equilibrium MD calculations \cite{LG05}. The small shear
rate data were found to obey a universal scaling: $\eta$ normalized by the
Einstein frequency was found to depend only on the reduced temperature (ratio
of the temperature to melting temperature). The high shear rate simulations
based on method 2 unambiguously demonstrated a non-Newtonian behavior of the
Yukawa liquid: $\eta$ was significantly reduced for these conditions in a
manifestation of shear thinning, except at the lowest shear rates where the
liquid is more nearly Newtonian. Regimes of the plasma coupling parameter were
identified to distinguish whether the kinetic or potential contribution to the
shear viscosity is primarily responsible for the shear thinning effect.

We thank G. J. Kalman and Bin Liu for useful discussions. This work
was supported by the Hungarian Fund for Scientific Research and the
Hungarian Academy of Sciences, OTKA-T-48389, MTA-OTKA-90/46140,
OTKA-PD-049991; J.G. was supported by NASA and DOE. We acknowledge
significant contributions from the referees of our paper, especially
for suggesting to use method 2 for our shear thinning calculations.

\begin{figure}
\caption{(a) Velocity profiles $v_x(y)$ obtained from method 1 for different
frequencies ($1/\tau$) of momentum exchange steps, and (b) $T_y(y)$ temperature
profiles for the same conditions. $\Gamma$ = 100, $\kappa$ = 1.
\label{fig:examples}}
\end{figure}

\begin{figure}
\caption{ (a) Trajectories of particles in the simulation based on
method 2. (a) $\Gamma$ = 10, $\kappa$ = 1, at a shear rate
$\overline{\gamma}$ = 0.2, time of recording : $\omega_{\rm p}
\Delta T$ = 5.0; (b) $\Gamma$ = 100, $\kappa$ = 1,
$\overline{\gamma}$ = 0.05, $\omega_{\rm p} \Delta T$ = 23.6. The
shear field is $v_x = \gamma (y-L_y/2)$, i.e. there is no flow at
$y = L_y/2$. $N$ = 1020. \label{fig:sllod}}
\end{figure}

\begin{figure}
\caption{ (a) Shear viscosity at near-equilibrium conditions, $\eta_{\rm
{eq}}$, obtained from method 1, at the simulation's lowest practical shear
rate, and normalized by $\eta_0 = m n \omega_{\rm p} a^2$. For comparison, data
are shown from the Iowa equilibrium MD simulation and from simulations based on
method 2 (SLLOD), in the limit of small shear rates, at $\kappa$ = 1. $N$ is
the number of simulation particles. (b) A scaling law is demonstrated by
normalizing the data in (a) using $\eta_{\rm E} = m n \omega_{\rm E} a^2$ and
$T'=T_y/T_{\rm m}$ where $T_m$ is the melting temperature. The thick line is an
empirical fit of form $\eta_{\rm {eq}} / \eta_{\rm E} = a T' + b/T' + c$.
\label{fig:results}}
\end{figure}

\begin{figure}
\caption{ (a) Shear viscosity as a function of normalized shear
rate $\overline{\gamma}$ for selected values of the coupling
parameter $\Gamma$. (b) Potential (filled symbols) and kinetic
(open symbols) contributions to the shear viscosity. $\kappa$ =
1.\label{fig:thinning}}
\end{figure}

\end{document}